\documentclass[10pt,preprint]{emulateapj}

\slugcomment{}

\shorttitle{Simulations for Terrestrial Planets Formation}
\shortauthors{Zhang N. and  Ji J.}

\begin{document}

\title{Simulations for Terrestrial Planets Formation}

\author{Niu ZHANG\altaffilmark{1,2}, Jianghui JI\altaffilmark{2,3}}
\email{jijh@pmo.ac.cn}

\altaffiltext{1}{Graduate School of Chinese Academy of Sciences,
Beijing 100049, China}

\altaffiltext{2}{Purple  Mountain  Observatory, Chinese  Academy of
Sciences,  Nanjing  210008, China}

\altaffiltext{3}{National Astronomical Observatory, Chinese Academy
of Sciences,Beijing 100012, China}

\begin{abstract}
In this paper, we investigate the formation of terrestrial planets
in the late stage of planetary formation using two-planet model. At
that time, the protostar has formed for about 3 Myr and the gas disk
has dissipated. In the model, the perturbations from Jupiter and
Saturn are considered. We also consider variations of the mass of
outer planet, and the initial eccentricities and inclinations of
embryos and planetesimals. Our results show that, terrestrial
planets are formed in 50 Myr, and the accretion rate is about $60\%
- 80\%$. In each simulation, 3 - 4 terrestrial planets are formed
inside "Jupiter" with masses of $0.15 - 3.6 M_{\oplus}$. In the 0.5
- 4 AU, when the eccentricities of planetesimals are excited,
planetesimals are able to accrete material from wide radial
direction. The plenty of water material of the terrestrial planet in
the Habitable Zone may be transferred from the farther places by
this mechanism. Accretion could also happen a few times between two
major planets only if the outer planet has a moderate mass and the
small terrestrial planet could survive at some resonances over time
scale of $10^8$ yr. In one of our simulations, com-mensurability of
the orbital periods of planets is very common. Moreover, a
librating-circulating 3:2 configuration of mean motion resonance is
found.
\end{abstract}

\keywords{astrophysics-exoplanet-planetary formation-n-body
simulation}

\section{Introduction}
Since the discovery of the first extrasolar planet around Solar-Type
star, the detection of the extrasolar planets develops rapidly. To
date, more than $300$ planets are found orbiting their center stars
beyond our solar system, including 35 multiple planetary systems.
Recently, scientists have found evidences of methane and carbon
dioxide in the atmosphere of a Hot-Jupiter (HD 189733 b)
(http://planetquest.jpl.nasa.gov). Of $\sim 300$ known extrasolar
planets, the minimum mass is generally several Jupiter masses. There
are also several terrestrial planets (Super-Earth), but their
orbital characteristic is unsuitable for the formation or
development of life. Along with the development of survey techniques
and incoming high definition space missions, people will definitely
discover more and more Earth-like planets in the extrasolar
planetary systems. The research of formation and evolution of the
terrestrial planet now becomes important topics in astrophysics,
astrobiology, astrochemistry and so on.

Planet formation has certain order \citep{zhou05}, and Jupiter-like
planets at greater distance are formed faster than those near the
Sun. It is generally believed that the planet formation may
experience the following stages: The grains condensed in the initial
stage grow to km-sized planetesimals in the early stage, and then,
in the middle stage, Moon-to-Mars sized embryos are formed by
accretion of the planetesimals. The size of embryos correlates with
the feeding zone of the planetesimals. According to the formula of
Hill radius: $R_{H} = r(m/3M_{\odot})^{1/3}$ (where $r$,$m$ are the
heliocentric distance and the mass of planetesimal), the distant
planetesimals have wider feeding zones, so the formed embryos are
larger. When the embryos grow to the core of mass ($\sim 10M_\oplus$
), runaway accretion may take place accordingly. With more
atmosphere accreted, the embryos contract, growing ever denser and
more massive, eventually collapse to form giant Jovian planets
\citep{hu08, ida04}. However, at the same period, the inner
planetesimals accrete in respective accretion scope, and then the
embryos of terrestrial planet (namely kind of the terrestrial planet
core matter) are formed. At the end of the third stage, it is around
that the protostar has formed for about $3$ Myr, the gas disk has
dissipated. A few larger bodies with low $e$ and $i$ are in crowds
of planetesimals with certain eccentricities $e$, and inclinations
$i$. In the late stage, the terrestrial planetary embryos are
excited to high eccentricity orbit by gravitational perturbation.
Then, the orbital crossing makes the planets accreting material in
the broader radial area. Solid residue is either scattered out of
the planetary system or accreted by the massive planet. However, it
also has the possibility of being captured at the resonance position
of the major planets \citep{nag00, hu08}.

Taking our Solar System as the background, \citet{cha01}
made a study of terrestrial planet formation in the late stage by
numerical simulations. He set $150 - 160$ Moon-to-Mars size
planetary embryos in the area of $0.3 - 2.0$ AU, include
gravitational perturbations from Jupiter and Saturn. He also
examined two initial mass distributions: approximately uniform
masses, and a bimodal mass distribution. The results show that $2 -
4$ planets are formed in $50$ Myr, and finally survive over $200$
Myr timescale. The space distribution and concentration (see Section
4 in \citet{cha01}) of planets formed in the simulations are similar
to our solar system. However, the planets produced by the
simulations usually have eccentric orbits with higher eccentricities
$e$, and inclinations $i$ than Venus.  \citet{ray04,
ray06} also studied the formation of terrestrial planets. In
the simulations, they took into account Jupiter's gravitational
perturbation, wider distribution of material ($0.5 - 4.5$ AU) and
higher resolution. The results confirm a leading hypothesis for the
origin of Earth's water: they may come from the material in the
outer area by impacts in the late stage of planet formation.
\citet{ray06b} explored the planet formation under planetary
migration of the giant. In the simulations, super Hot Earth form
interior to the migrating giant planet, and water-rich, Earth-size
terrestrial planet are present in the Habitable Zone ($0.8 - 1.5$
AU) and can survive over $10^8$ yr timescale.

In our model, Solar System is taken as the background. But several
changes are worth noting : 1) we use two-planet (Jupiter and Saturn,
see Fig.1) model. 2) in the model, Jupiter and Saturn are supposed
to be formed at the beginning of the simulation, with two swarms of
planetesimals distributed among $0.5 - 4.2$ AU and $6.2 - 9.6$ AU
respectively. 3) The initial eccentricities and inclinations of
planetesimals are considered. 4) The variations of the mass of
Saturn are examined. 5) The exchange of material in the radial
direction is also studied by the parameter of water mass fraction.
6) We perform the simulations over longer timescale ($400$ Myr) in
order to check the stability and the dynamical structure evolution
of the system. Our results show that the terrestrial planets
produced interior to Jupiter have higher mass accretion rate, and
share the similar architecture as the Solar System. However, the
structure beyond Jupiter correlates with the initial mass of Saturn.
Almost each simulation has a water-rich terrestrial planet in the
Habitable Zone ($0.8 - 1.5$ AU).

In Section 2, the initial conditions, algorithm and integration
procedure are described in detail. Section 3 presents the main
results. We conclude the outcomes in Section 4.

\section{Model}
\subsection{Initial conditions}
Generally, the time scale for formation of Jupiter-like planet is
less than $10$ Myr \citep{bri01}. Nevertheless, as we know, the
formation scenario of planet embryos is related to their
heliocentric distances and the initial mass of the star nebular. If
we consider the model of $1.5$ MMSN (minimum mass solar nebular),
the upper bound of the time scale for Jupiter-like planet formation
corresponds to the time scale for the embryo formation at $2.5$ AU
\citep{kok02}, which is just at $3:1$ resonance location of Jupiter.
In the region $2.5 - 4.2$ AU, embryos will be cleared off by strong
gravitational perturbation arising from Jupiter. There should be
some much smaller solid residue among Jupiter and Saturn, even
though the 'clearing effect' may throw out most of the material in
this area. That's why we set embryos only in the region $0.5 - 2.5$
AU and planetesimals in the $0.5 - 4.2$ AU and $6.2 - 9.6$ AU.

We adopt the surface density profile as follows \citep{ray04}:
\begin{equation}
\Sigma (r) = \left\{
\begin{array}{ll}
\Sigma_{1}r^{-3/2}, & r < snow~line,\\
\Sigma_{snow}(\frac{r}{5AU})^{-3/2}, & r > snow~line.
\end{array}
\right.
\end{equation}

In (1), $\Sigma_{snow} = 10~g/cm^2 $ is the surface density at
snowline, the snowline is at $2.5$ AU with $\Sigma_1 = 4~g/cm^2 $.As
mentioned earlier, the mass of planetary embryos is proportional to
the width of the feeding zone, which is associated with Hill Radius,
$R_{H} $, so the mass of an embryo increases as
\begin{equation}
M_{embryo} \propto r\Sigma(r) R_{H}
\end{equation}

The embryos among $0.5 - 2.5$ AU are spaced by $\Lambda $ ($\Lambda
$ varying randomly between 2 and 5) mutual Hill Radii, $R_{H,m}$ ,
which is defined as
\begin{equation}
R_{H,m} = (\frac{a_1+a_2}{2})(\frac{m_1+m_2}{3M_{\odot}})^{1/3}
\end{equation}
where $a_{1,2} $ and $m_{1,2} $ are the semi-major axes and masses
of the embryos respectively. Replacing $R_{H} $ in (2) with $R_{H,m}
$ , and substituting (1) in (2), then, we get relations between the
mass of embryos and the parameter $\Lambda $ as
\begin{equation}
M_{embryo} \propto r\Sigma(r) R_{H,m} \propto
r^{3/4}\Lambda^{3/2}\Sigma^{3/2}
\end{equation}

As shown in Fig. 1, the initial planetesimals are spread over $0.5 -
9.6$ AU (excluding $3$ Hill Radii around Jupiter); the distribution
of them should meet (1). Here, we equally set the masses of
planetesimals inside and outside Jupiter respectively as shown in
(5). Consequently, the number distribution of the planetesimals is
only needed to satisfy $N\propto r^{-1/2} $. Additionally, we keep
the total number of planetesimals and embryos inside Jupiter, and
the number of planetesimals outside Jupiter both equal to $200$.
\begin{equation}
\left\{
\begin{array}{ll}
\sum N_{embryo} + \sum N_{planetesimal,~~r < r_{Jupiter}} \\
~~~~= \sum N_{planetesimal,~r > r_{Jupiter}} = 200,\\
\sum M_{embryo} + \sum M_{planetesimal,~r < r_{Jupiter}} \\
~~~~= \sum M_{planetesimal,~r > r_{Jupiter}} = 7.5M_{\oplus},\\
M_{planetesimal,~r < r_{Jupiter}} = \\
~~~~ (7.5M_{\oplus} -
\sum M_{embryo}) / (200 - \sum N_{embryo}),\\
M_{planetesimal,~r > r_{Jupiter}} = 7.5M_{\oplus} / 200.
\end{array}
\right.
\end{equation}

The water mass fraction of the bodies is same as \citet{ray04},
i.e., the planetesimals beyond 2.5 AU have $5\%$ water material by
mass, those between $2 - 2.5$ AU have $0.1\%$ water material by
mass, and the others have $0.001\%$ water material by mass. The
eccentricities and inclinations vary in ($0 - 0.02$) and ($0 -
0.05^\circ$), respectively. The mass of Saturn in simulations
1a/1b,2a/2b and 3a/3b are $0.5M_\oplus $, $5M_\oplus $, $50M_\oplus
$ respectively. Each simulation is carried out twice with a)
considering, b) not considering self-gravitation of planetesimals
among Jupiter and Saturn.

\subsection{Algorithm}
In regular coordinate system, the motion equations of an n-body
system are \citep{murbook99}

\begin{equation}
\left\{ \begin{array}{ll}
\frac{dx_i}{dt} &= \frac{\partial H}{\partial p_i}, \\
\frac{dp_i}{dt} &= - \frac{\partial H}{\partial x_i}.
\end{array} \right.
\end{equation}
where the index $i=1,2,\cdots,n $ denotes the body $i$ , and $x_i$ ,
$p_i $ are the generalized coordinate and momentum of the body $i$,
respectively. The Hamiltonian,
$H=\sum\limits_{i=1}^{n}\frac{p_i^2}{2m_i}-G\sum\limits_{i=1}^{n}m_i
\sum\limits_{j=i+1}^{n}\frac{m_j}{r_{ij}}$, is the sum of the
kinetic and potential energy for the system. From (6), we know that
the rate of any quantity, $q$ , can be conveniently expressed in the
following form,
\begin{equation}
\begin{array}{ll}
\frac{dq}{dt}&=\sum\limits_{i=1}^{n}(\frac{\partial q}{\partial x_i}
\frac{dx_i}{dt}+\frac{\partial q}{\partial p_i}\frac{dp_i}{dt})\\
&=\sum\limits_{i=1}^{n}(\frac{\partial q}{\partial x_i}
\frac{\partial H}{\partial p_i}-\frac{\partial q}{\partial p_i}
\frac{\partial H}{\partial x_i}).
\end{array}
\end{equation}
If we define an operator $F=\sum\limits_{i=1}^{n}(\frac{\partial
~}{\partial x_i}\frac{\partial H}{\partial p_i}-\frac{\partial
~}{\partial p_i}\frac{\partial H}{\partial x_i}) $ \citep{cha99},
then we can rewrite (7) as $\frac{dq}{dt}=Fq $. Integral the
differential equation over time $t_1-t_2 $ ($t_2 > t_1 $), and so we
get
\begin{equation}
q_2=e^{(t_2-t_1)F}q_1,
\end{equation}
where $q_1$ and $q_2$ are the values of $q$ corresponding to the
time $t_1$ and $t_2$ respectively. If we define $h=t_2-t_1 $ ($h$
actually is the internal time step), then expand (8) at zero, we
have
\begin{equation}
q_2=(1+hF+\frac{h^2F^2}{2}+\cdots)q_1,
\end{equation}
The symplectic integrator is to divide $H$ into pieces, each piece
could be individually solved, and then they approximate the solution
of the problem via applying the solutions once a time. For example,
we split the Hamiltonian $H=H_1+H_2 $ , and hence have the operators
$F_1,F_2 $ . It is easy to obtain $q_2=e^{h(F_1+F_2)}q_1 $ from (8).
By expanding the exponential (attn. $F_1F_2 \neq F_2F_1 $), ignoring
the second- and higher-order small quantities of $h$ , then we have
\begin{equation}\begin{array}{ll}
e^{h(F_1+F_2)} & =e^{hF_1}e^{hF_2}+\frac{h^2(F_2F_1-F_1F_2)}{2}\\
& =1+h(F_1+F_2)+\frac{h^2(F_1^2+2F_1F_2+F_2^2)}{2}+\cdots
\end{array}
\end{equation}
We can get a second-order integrator
$q_2=e^{hF_2/2}e^{hF_1}e^{hF_2/2}q_1 $ by applying a small
equivalence transformation.

The key point for symplectic algorithm is how to split Hamiltonian
$H$ into pieces. Considering a dynamical system composed of $N$
bodies orbiting a massive central body, we can split the Hamiltonian
$H$ into the primary and the secondary parts. \citet{cha99} proposed
a hybrid symplectic algorithm, in which the Hamiltonian   is divided
into the following parts:
\begin{equation}
\left\{\begin{array}{ll}
H_1&=\sum\limits_{i=1}^{N}(\frac{p_i^2}{2m_i}-\frac{Gm_{\odot}m_i}{r_{i\odot}}),\\
H_2&=-G\sum\limits_{i=1}^{N}\sum\limits_{j=i+1}^{N}\frac{m_im_j}{r_{ij}},\\
H_3&=\frac{1}{2m_{\odot}}(\sum\limits_{i=1}^{N}{\bf \it p_i})^2,
\end{array}
\right.
\end{equation}
where $H_1 $ is the unperturbed Keplerian motion of the $N$ smaller
bodies, $H_2 $ is the total interaction potential energy of the $N$
smaller bodies, and $H_3 $ is the kinetic energy of the center body
(Note That: $N$ has different meaning from $n$ above, $N$ refers to
the numbers of bodies excluding the central body). The term of
'hybrid' means that, for the convenience of calculation,
heliocentric coordinates and barycentric velocities are used while
solving (11) \citep{cha99, dun98}. From (10), we can infer that all
of the higher terms depend on both $F_1$ and $F_1$. If $F_2\sim
\epsilon F_1 $ ($\epsilon=\sum m_i/m_{\odot} $), and therefore, the
second-order integrator
$q_2=e^{hF_2/2}e^{hF_3/2}e^{hF_1}e^{hF_3/2}e^{hF_2/2}q_1 $ is
correct to $O(\epsilon h^3) $ only when the different parts of
Hamiltonian meet the conditions $H_1\gg H_2,H_1\gg H_3 $. However,
when close encounter occurs between two bodies, the distance between
them $r_{ij} $ approaches zero, hence the $H_1\gg H_2 $ can not be
satisfied. \citet{cha99} introduced a changeover function $K(r_{ij})
$ to translate part of $H_2 $ associated with the close encounter to
$H_1 $, and then integrate it using Bulirsch-Stoer method
\citep{stobook80}. The modified $H_1,H_2 $ are present as
\begin{equation}
\left\{\begin{array}{ll}
H_1&=\sum\limits_{i=1}^{N}(\frac{p_i^2}{2m_i}-\frac{Gm_{\odot}m_i}{r_{i\odot}})\\
&~~~~-G\sum\limits_{i=1}^{N}\sum\limits_{j=i+1}^{N}\frac{m_im_j}{r_{ij}}[1-K(r_{ij})],\\
H_2&=-G\sum\limits_{i=1}^{N}\sum\limits_{j=i+1}^{N}\frac{m_im_j}{r_{ij}}K(r_{ij}).
\end{array}
\right.
\end{equation}
$K(r_{ij}) $ tends to zero when $r_{ij} $ is small, while tending to
one when $r_{ij} $ is large \citep{cha99}.

We use the hybrid symplectic integrator \citep{cha99} in MERCURY
package to integrate all the simulations. We take into account that
collision and coalescence will occur, when the minimum distance
between any of the two objects is equal to or less than the sum of
their physical radii. While they were separated by not more than 3
Hill radii, we consider close encounters will take place. When the
distance from the central star is more than $100$ AU, these bodies
are removed, because they are so far from the central star that they
play an insignificant role of the interaction. In addition, we adopt
$6$ days as the length of time step, which is a twentieth period of
the innermost body at $0.5$ AU. The $6$ simulations are carried out
over $400$ Myr time scale. At the end of the intergration, the
changes of energy and angular momenta are $10^{-3}$ and $10^{-11}$
respectively. The $6$ simulations are performed on a workstation
composed of $12$ CPUs with $1.2$ GHz, and each costs roughly $45$
days.

\section{Results}
All of $6$ simulations exhibit some classical processes on planet
formation. Firstly, we will analyse simulation 2a/2b to discuss the
physical processes which can apply to every simulation. Next, we
will make a statistical analysis in order to find out that how the
planet formation may rely on different physical factors.

\subsection{Simulation 2a/2b}
At the end of the calculation, $3-4$ terrestrial planets are formed
in 2a/2b. Table 1 shows the properties of the terrestrial planets
from simulations 2a and 2b. We label the planets as b, c, d, and so
on, according to the heliocentric distance (hereinafter). The masses
of the terrestrial planets range from several Mars masses to several
Earth masses. All of them are water-rich, except the planet e in
simulation 2b. Some parameters of a certain planet are comparable
with the terrestrial planets in solar system. For example, the
orbital eccentricity of planet b in simulation 2b is $0.0309$, which
is very close to that of Earth.

Fig. 2 is a snapshot of simulation 2a. At $0.1$ Myr, it is clear
that the planetesimals are excited at the $3:2$ ($3.97$ AU),$2:1$
($3.28$ AU) and $3:1$ ($2.5$ AU) resonance locations with Jupiter,
and this is similar to the Kirkwood gaps of the asteroidal belt in
solar system. For about $1$ Myr, planetesimals and embryos are
deeply intermixed, most of the bodies have large eccentricities.
Collisions and accretions emerge among planetesimals and embryos.
This process continues until about $50$ Myr, the planetary embryos
are mostly formed, and then dynamical evolution is start. The
formation time scale in our work is in accordance with that of
\citep{ida04}. Finally, inside Jupiter, $3$ terrestrial planets are
formed with masses of $0.15 - 3.6 M_{\oplus}$. However, at the outer
region, planetesimals are continuously scattered out of the system
at $0.1$ Myr. For about $10$ Myr, there are no survivals except at
some resonances with the giant planet. As shown in Fig. 2, there is
a small body at the $1:2$ resonance with Jupiter. Due to the
planetesimals' scattering, Jupiter (Saturn) migrates inward
(outward) $0.13$ AU ($1.19$ AU) toward the sun respectively. Such
kind of migration agrees with the work of Fernandez et al.
\citep{fer84} Hence, the $2:5$ mean motion resonance is destroyed,
then the ratios of periods between Jupiter and Saturn degenerate to
$1:3$. Therefore, the ratio of periods for Jupiter, small body and
Saturn is approximate to $1:2:3$.

Fig. 3 is a snapshot of simulation 2b. In comparison with Fig. 2, it
is apparent that planetesimals are excited more quickly at the $3:2$
($3.97$ AU), $2:1$ ($3.28$ AU) and $3:1$ ($2.5$ AU) resonance
location with Jupiter. The several characteristic time scales are
the same as simulation 2a for the bodies within Jupiter. 4 planets
are formed in simulation 2b, the changes of position of Jupiter and
Sat-urn are about the same as simulation 2a. We note simulations 2a
and 2b have the same initial conditions, the only difference between
them is whether we consider the self-gravitation among the outer
planetesimals. The results of simulation 2b are shown to be a
consequence of being expected but not surprising. There is a little
stack planetesimals survival over $400$ Myr among $7 - 8$ AU,
located in the area of $2:3$ ($6.63$ AU) and $1:2$ ($8.03$ AU)
resonances with Jupiter.

Planets in 55 cnc planetary system have similar spatial distribution
to the solar system \citep{fis08}, From Table 1 and Fig. 3, it is
not difficult to see that 4 terrestrial planets formed in simulation
2b move on the nearly-circular orbit. Mars is ever regarded as a
survivor of an original planetary embryo, according to its unique
chemical and isotopic characteristics. As a matter of fact, the
planet e in simulation 2b is a survivor of the initial planetary
embryos. Planet e in simulation 2b does not accrete anything over
$400$ Myr integration time. Comparing the semi-major axis of Mars
with that of planet e, we notice that the planetesimals of
simulation 2b are located in the asteroidal belt. Furthermore, the
ratio of periods of the planet e and Jupiter is nearly $1:2$. The
total mass of the main-belt in solar system is about $5\times
10^{-5}M_{\oplus} $, being $0.1\%-0.12\% $ \citep{hu08} of the
initial solid material. If the assumed planet e in simulation 2b
would break into thousands of fragments, they may undergo
re-accretion or ejection by the perturbation of Jupiter over secular
evolution. If it is similar to the same ratio of mass of the belt in
solar system, then the leftovers of the solid materials almost bear
a total mass of $6.3\times 10^{-5}-7.56\times 10^{-5}M_{\oplus} $ .
In this sense, an asteroidal belt is very likely to form in the
system quite similar to that of our solar system.

Fig. 4a is the mass curve of terrestrial planets for simulation 2b.
We can find that the accrete velocity is not uniform. At $10$ Myr,
planets reach half mass of their final mass, and then, the accretion
velocity slows down, because the planetesimals are only a quarter
left. Until about $50$ Myr, the terrestrial planets are formed. The
accretion rate and mass concentration (the mass rate of the largest
terrestrial planet and the total formed objects) are $73\%$ and
$43\%$, respectively. The corresponding parameters in simulation 2a
are $60\%$ and $81\%$ respectively. However, in the area outside
Jupiter, $81\%$ initial material is scattered out of the system.

Planet embryos are formed from feeding zones where the planetesimals
are located. A feeding zone has unique chemical and isotopic
characteristics. It is helpful to study the trace of the
planetesimals to understand the composition of terrestrial planet,
vice versa, for example, if we can investigate the origin and
formation process by revealing the chemical or isotopic
characteristics of the moon. In Fig. 4b is shown the trace of
survivals. We note that all the materials accreted by terrestrial
planets come from the inner swarm of planetesimals or embryos. Here
Jupiter is like a wall, which separates the inner and outer
planetesimals from exchanging materials. Once again, Fig. 4b
verifies that Jupiter may protect the inner terrestrial planets from
colliding with the outer bodies \citep{wet90}. We set a water mass
fraction on each body, it is easy to work out how much
water-material of the finally terrestrial planet bears. Take planet
c in simulation 2b for example, the water material is approximately
$1.1\times10^{22}kg $ , about 8 times than Earth. Fig. 4b can also
verify that terrestrial planet accrete material in broad radial
direction.

\subsection{Statistical analysis}
The production efficiency of the terrestrial planet in our model is
high, and the accretion rate inside Jupiter is $60\% - 80\%$ in the
simulations. $3 - 4$ terrestrial planets formed in $50$ Myr. 5 of 6
simulations have a terrestrial planet in the Habitable Zone ($0.8 -
1.5$ AU) (see Fig. 5). The planetary systems are formed to have
nearly circular orbit and coplanarity, similar to the solar system
(see Table 2). We suppose that the above characteristics are
correlated with the initial small eccentricities and inclinations.
Such adoption could generate more close encounters or collisions in
the early several Myr, which may increase viscosity of the system
and then make the orbits more nested on circular orbit on the
orbital plane. The concentration in Table 2 means the ratio of
maximum terrestrial planet formed in the simulation and the total
terrestrial planets mass. It represents different capability on
accretion, and is not associated with self-gravitation. The average
value of this parameter is similar to the solar system. Considering
the self-gravitation of planetesimals among Jupiter and Saturn, the
system has a better viscosity, so that the planetesimals will be
excited slower. The consideration of self-gravitation may not change
the formation time scale of terrestrial planets, but will affect the
initial accretion speed and the eventual accretion rate. In Table 2,
the simulations 1b, 2b, 3b have a bit higher accretion rate. When
the self-gravitation is not considered, the planetesimals may be
excited quickly. The accretion has a faster speed at the early
several Myr, so this can promote the accretion rate.

From Fig. 4b, we have to be aware of that Saturn accretes a few
planetesimals, this is uncommon in simulations 1a, 1b, 3a, 3b. And
Fig. 5 shows the finally structure of the simulations, it is clear
that different Saturn mass will affect the outer structure of the
system beyond Jupiter. In simulation 3a (3b), the Saturn mass is
$50M_{\oplus} $ . Now it is large enough to clear the area among
Jupiter and Saturn. In simulation 1a (1b), Saturn's mass is
$0.5M_{\oplus} $ , more or less equal to the embryos' mass. In this
case, it is too small to clear off any planetesimal amongst the
region of Jupiter and Saturn. Therefore, we draw the conclusion that
only the Saturn's mass is close to be $5-10M_{\oplus}$ , then
accretion may happen. Saturn and Jupiter in our solar system may
form in same stage. If there exist embryos of $10M_{\oplus} $
outside Saturn, the giant Jupiter-mass planets may form.

The scattering of planetesimals could cause the migration of
planets, for example, Jupiter migrated from $5.2$ AU to $5.06$ AU
while Saturn traveled from $9.6$ AU to $10.71$ AU. There are plenty
of ratios of semi-major axis of the survival planets nearly $2:1$.
In this case, the planet is easily to be captured on $2:1$ mean
motion resonance \citep{lee02, lee04, zhou05, fis08}. There are
still some ratios of periods between the survival bodies close to a
simple ratio of integers (see Fig. 5). In the very long process of
dynamical evolution after planetary formation, the planets also have
the possibility of been captured onto resonant orbit. For example,
the orbits show the $2:3$ mean motion resonance be-tween Jupiter and
outer small body in simulation 1b, and the $3:1$ resonance between
Jupiter and the planet d in simulation 3a and so on. Many
researchers have studied the resonance and stability of the
planetary systems \citep{ji03, zhou03, zhou04}. As shown in Fig. 6,
in simulation 2a, two planets are on crossing orbits. When a close
encounter occurs, $3:2$ mean motion resonance is formed, with
resonance angle $\theta_1=2\lambda_1-3\lambda_2+\varpi_2 $ (where
$\lambda _{1,2} $ are the mean longitude and the longitudes of
periapse, the footnotes 1, 2 means the inner and outer planets
respectively.) librating around $180^\circ$ , while
$\theta_2=2\lambda_1-3\lambda_2+\varpi_1 $ circulating, and $e_1$
shows large oscillations. Such 'librating-circulating' configuration
is similar to the configuration of $2:1$ resonance in HD 73526
planetary system. There have been several hypotheses about its
origin \citep{tin06, san06, san07}. However, it still needs further
study in the future.

\section{Conclusions}
We simulate the terrestrial planets formation by using two-planet
model. In the simulation, the variations of the mass of outer
planet, the initial eccentricities and inclinations of embryos and
planetesimals are also considered. The results show that, during the
terrestrial planets formation, planets can accrete material from
different regions inside Jupiter. Among $0.5 - 4.2$ AU, the
accretion rate of terrestrial planet is $60\% - 80\%$, i.e., about
$20\% - 40\%$ initial mass is removed during the progress. The
planetesimals will improve the efficiency of accretion rate for
certain initial eccentricities and inclinations, and this also makes
the newly-born terrestrial planets have lower orbital
eccentricities. It is maybe a common phenomenon in the planet
formation that the water-rich terrestrial planet is formed in the
Habitable Zone. The structure, which is similar to that of solar
system, may explain the results of disintegration of a terrestrial
planet. Most of the planetesimals among Jupiter and Saturn are
scattered out of the planetary systems, and this migration caused by
scattering \citep{fer84} or long-term orbital evolution can make
planets capture at some mean motion resonance location. Accretion
could also happen a few times between two planets if the outer
planet has a moderate mass, and the small terrestrial planet could
survive at some resonances over $10^8$ yr time scale. Structurally,
Saturn has little effect on the architecture inside Jupiter, owing
to its protection. However, obviously, a different Saturn mass could
play a vital role of the structure outer Jupiter. Jupiter and Saturn
in the solar system may form over the same period.

In our simulations, neither terrestrial planets are formed within
$0.1$ AU, nor planetesimals or embryos are left. However, a lot of
exoplanets with orbital semi-major less than $0.1$ AU are observed,
and several Super-Earths are discovered. It is usually believed that
they were formed far from the center star and then migrated into
current location \citep{ray06b}. We do not consider the migration in
the simulations, which is caused by the interaction between the
giant planets or planetesimals in the gaseous disk \citep{ou07}. So
the simulation in this work can be applied to the case of the
dissipation of gas disk, in the late stage of planet formation. In
the future study, we will consider the giant planets under inward
migration, and in such circumstances short-period terrestrial planet
could be produced. To date, terrestrial planets are not detected in
the observations, due to the reasons of the selection effect of
detection methods and the low resolution precision. Both Doppler
velocities and transit method are sensitive to the objects moving in
smaller orbits. The current research of extrasolar terrestrial
planets has greatly contributed to the origin and evolution of our
own solar system. Kepler has been launched successfully on March 6,
2009, whose main scientific objective is detecting the Earth-like
terrestrial planets. Along with high accuracy incoming space
projects, it is predictable that more and more extrasolar planetary
systems with similar structure to the solar system will be
discovered.

\acknowledgments We are very grateful to Prof. Jilin Zhou of Nanjing
University for reading carefully and giving valuable suggestion to
improve the manuscript. We thank Prof. Qinglin Zhou and Dr.
Xiaosheng Wan of Ministry of Education Key Modern Astronomy and
Astrophysics Laboratory of Nanjing University for their kind help.
This work is financially supported by the National Natural Science
Foundation of China (Grants 10573040, 10673006, 10833001, 10233020)
and the Foundation of Minor Planets of Purple Mountain Observatory.

\clearpage

\begin{deluxetable}{rccccc}
\tablewidth{0pt} \tablecaption{Properties of terrestrial planets
from simulations 2a and 2b.} \tablewidth{0pt} \tablehead{
\colhead{Planet} & \colhead{$a$ (AU)} & \colhead{$e$} & \colhead{$i
(deg)$} & \colhead{$m (m_\oplus)$} & \colhead{water mass}}
\startdata
2a b & 0.6174 & 0.1098 &  7.452 & 3.6421 & 0.0316\% \\
   c & 1.7796 & 0.1935 & 33.890 & 0.1528 & 0.1040\% \\
   d & 2.3304 & 0.3291 &  9.393 & 0.6925 & 0.5218\% \\
\hline \\
2b b & 0.5274 & 0.0309 &  4.041 & 2.3527 & 0.1539\% \\
   c & 1.0451 & 0.1080 &  4.915 & 1.3880 & 0.1316\% \\
   d & 1.4783 & 0.0520 &  5.189 & 1.6696 & 0.7795\% \\
   e & 3.1162 & 0.2086 &  6.421 & 0.0630 & 0.0010\%
\enddata
\end{deluxetable}

\begin{deluxetable}{rcccccc}
\tablewidth{0pt} \tablecaption{Properties of terrestrial planets
from different systems} \tablewidth{0pt} \tablehead{
\colhead{System} & \colhead{accretion rate} & \colhead{$n$} &
\colhead{$\bar{m} (m_\oplus)$} & \colhead{concentration} &
\colhead{$\bar{e}$} & \colhead{$\bar{i} (^\circ)$}} \startdata
   1a & 73.2518\%    & 3   & 1.8313 & 0.4606 & 0.1381 &  7.6963 \\
   1b & 80.3853\%    & 3   & 2.0096 & 0.4262 & 0.0937 &  1.7790 \\
   2a & 59.8322\%    & 3   & 1.4958 & 0.8116 & 0.2108 & 16.9117 \\
   2b & 72.9779\%    & 4   & 1.3683 & 0.4299 & 0.0999 &  5.1415 \\
   3a & 65.1098\%    & 3   & 1.6277 & 0.5337 & 0.2063 &  5.9153 \\
   3b & 66.9694\%    & 3   & 1.6742 & 0.5040 & 0.1839 &  5.2447 \\
1a-3b & 69.7544\%    & 3.2 & 1.6678 & 0.5276 & 0.1554 &  7.1148 \\
solar &  -           & 4   & 0.4943 & 0.5058 & 0.0764 &  3.0624
\enddata
\end{deluxetable}
\clearpage

\begin{figure}
\figurenum{1}
\plotone{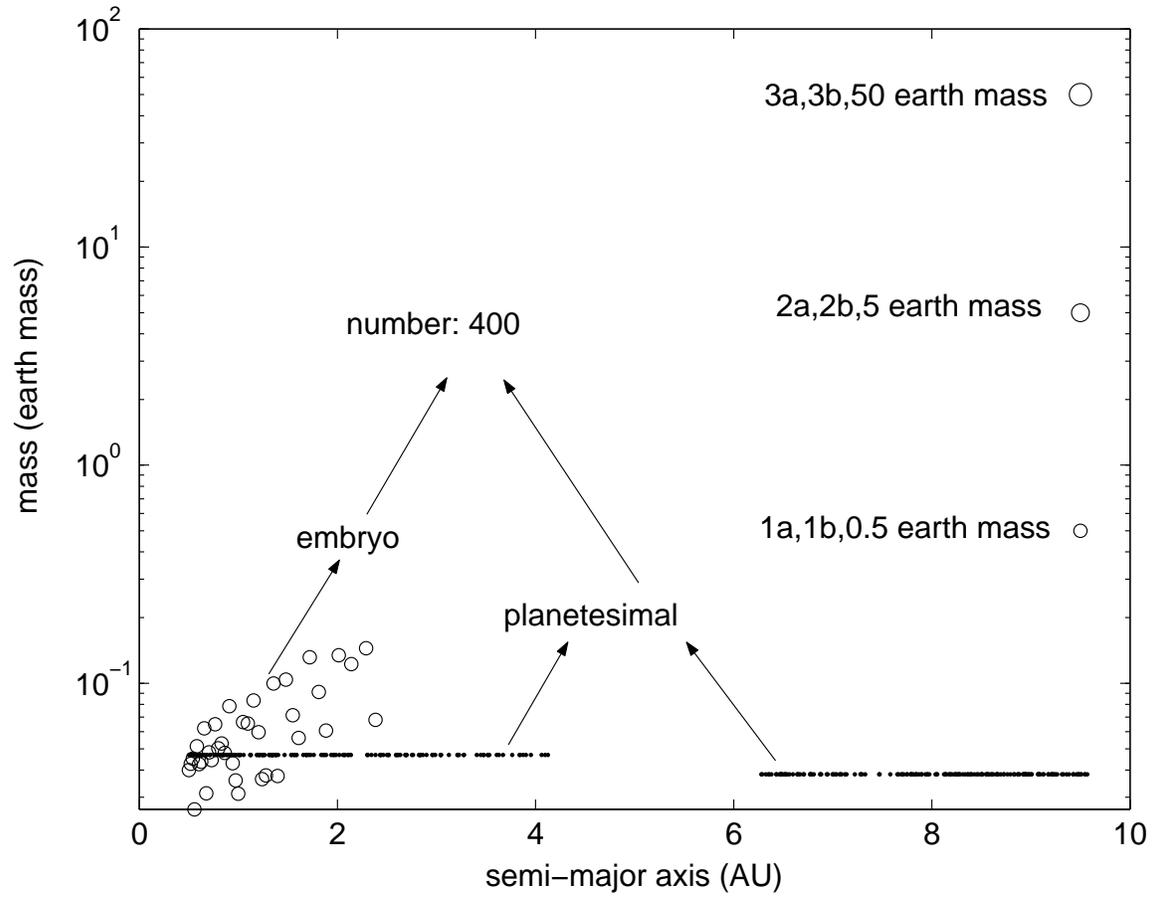}
\caption{Two-planet model and initial conditions.
\label{fig1}}
\end{figure}

\begin{figure}
\figurenum{2}
\plotone{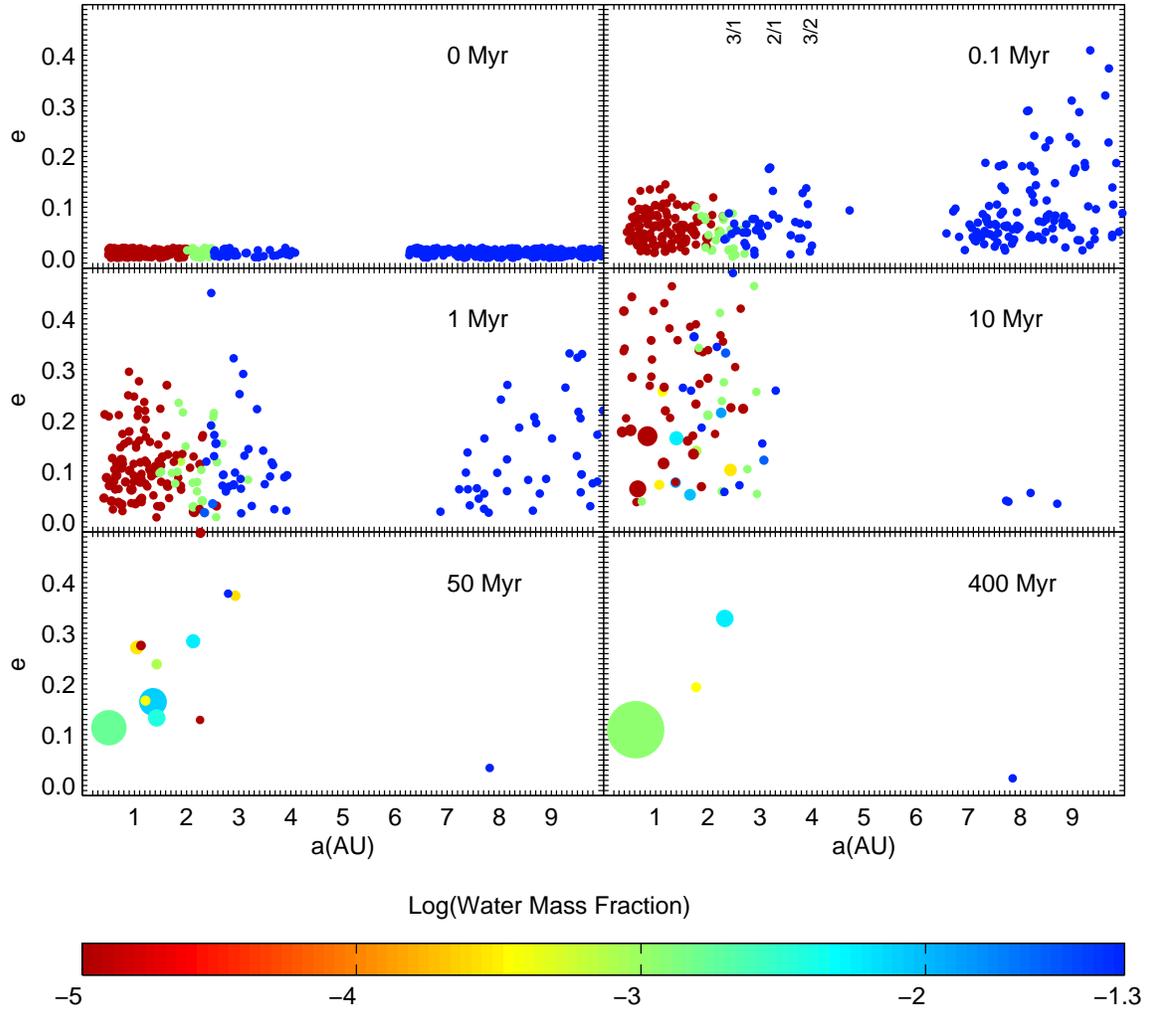} \caption{Snapshot of simulation 2a
with $M_{Saturn}=5M_{\oplus}$. The total mass of embryos is
$2.4M_{\oplus}$, the masses of planetesimals inside Jupiter are
$0.0317M_{\oplus}$, and those outside Jupiter are $0.0375M_{\oplus}
$. Planetesimals among Jupiter and Saturn were nonself-gravitational
(see Section 2.1). Note the size of each object is relative, and the
value bar is log of water mass fraction, e.g. the wettest body has
water mass fraction $\log _{10}(5\%)=-1.3$.
\label{fig2}}
\end{figure}

\begin{figure}
\figurenum{3}
\plotone{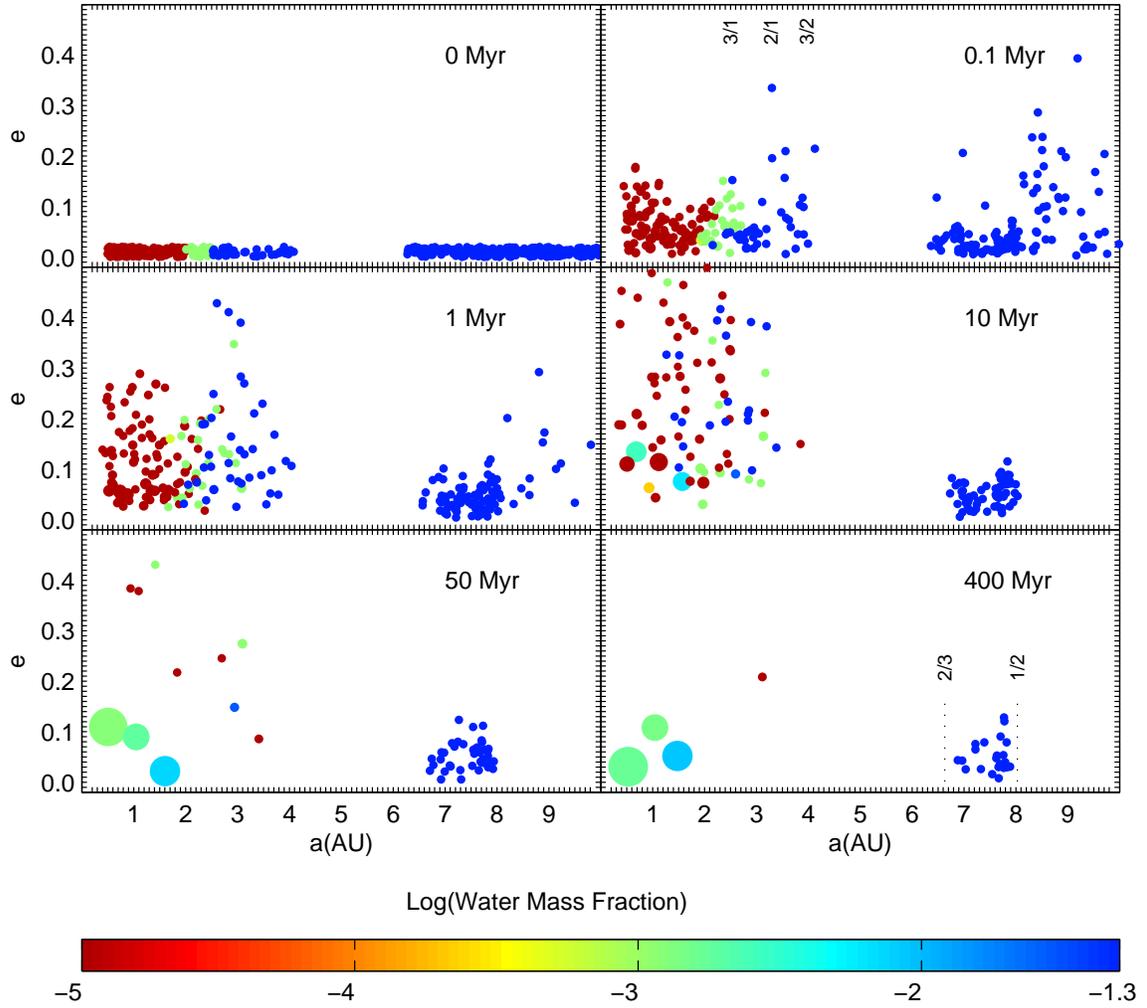}
\caption{Snapshot of simulation 2b.
\label{fig3}}
\end{figure}

\begin{figure}
\figurenum{4}
\plotone{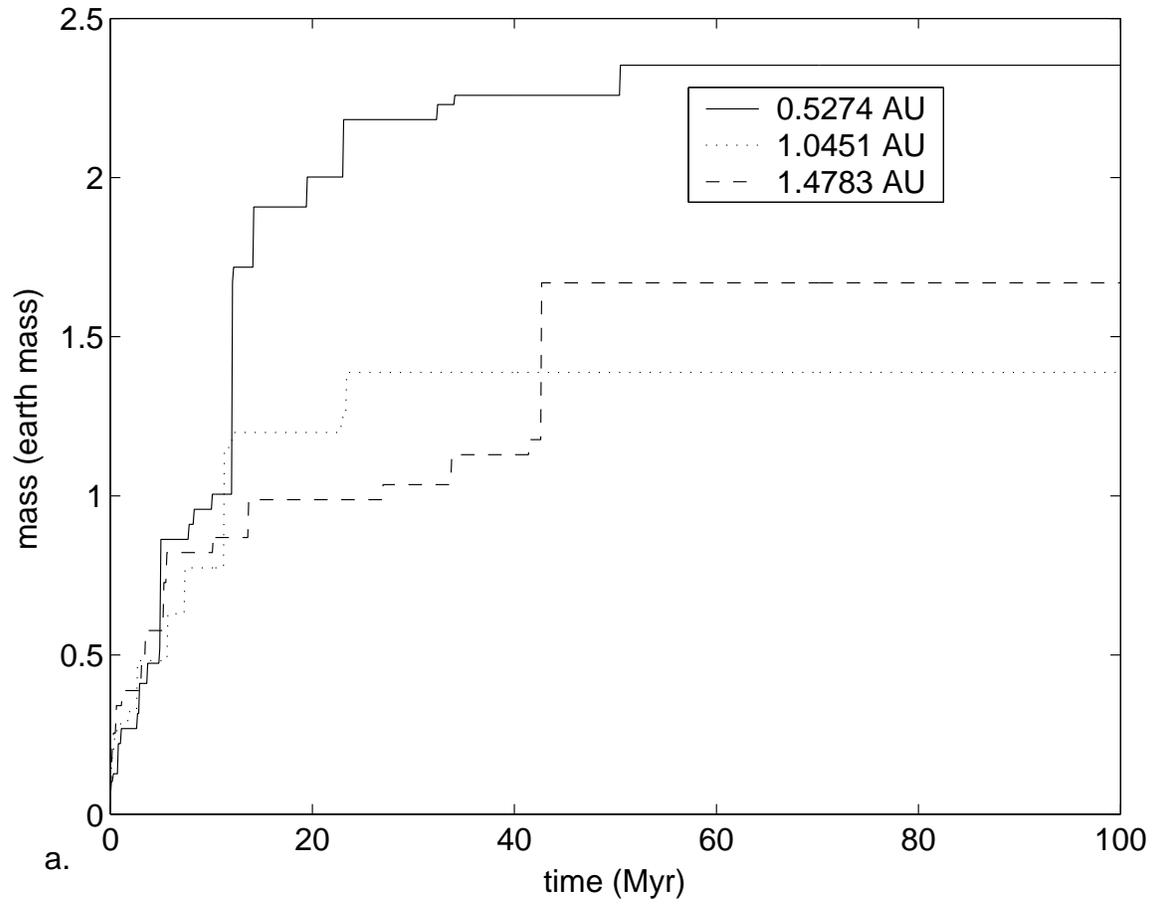}\caption{(a) Mass curve in simulation 2b.
\label{fig4}}
\end{figure}

\begin{figure}
\figurenum{4}
\plotone{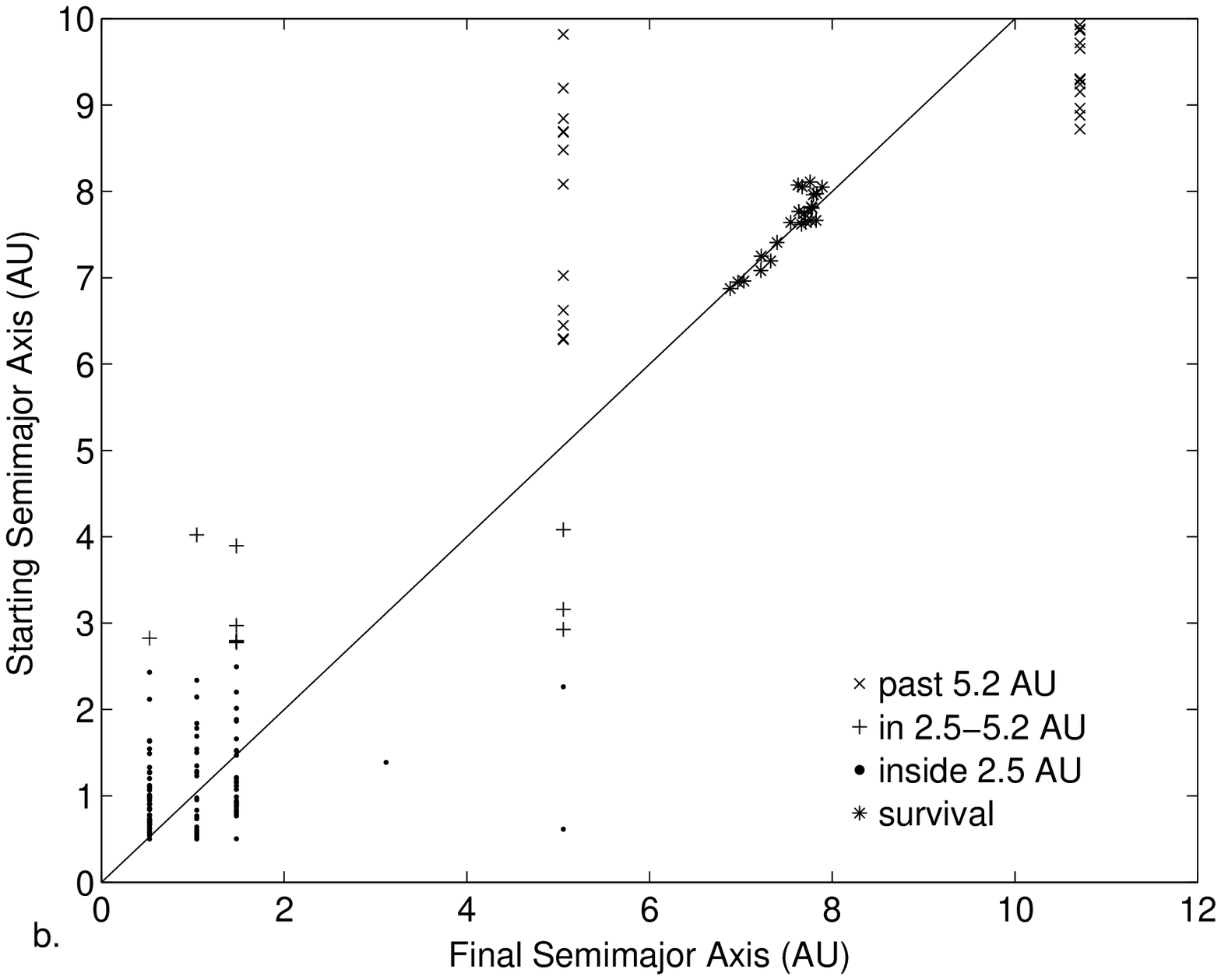}\caption{(b) Trace of the objects' in simulation 2b.
\label{fig4}}
\end{figure}

\begin{figure}
\figurenum{5} \plotone{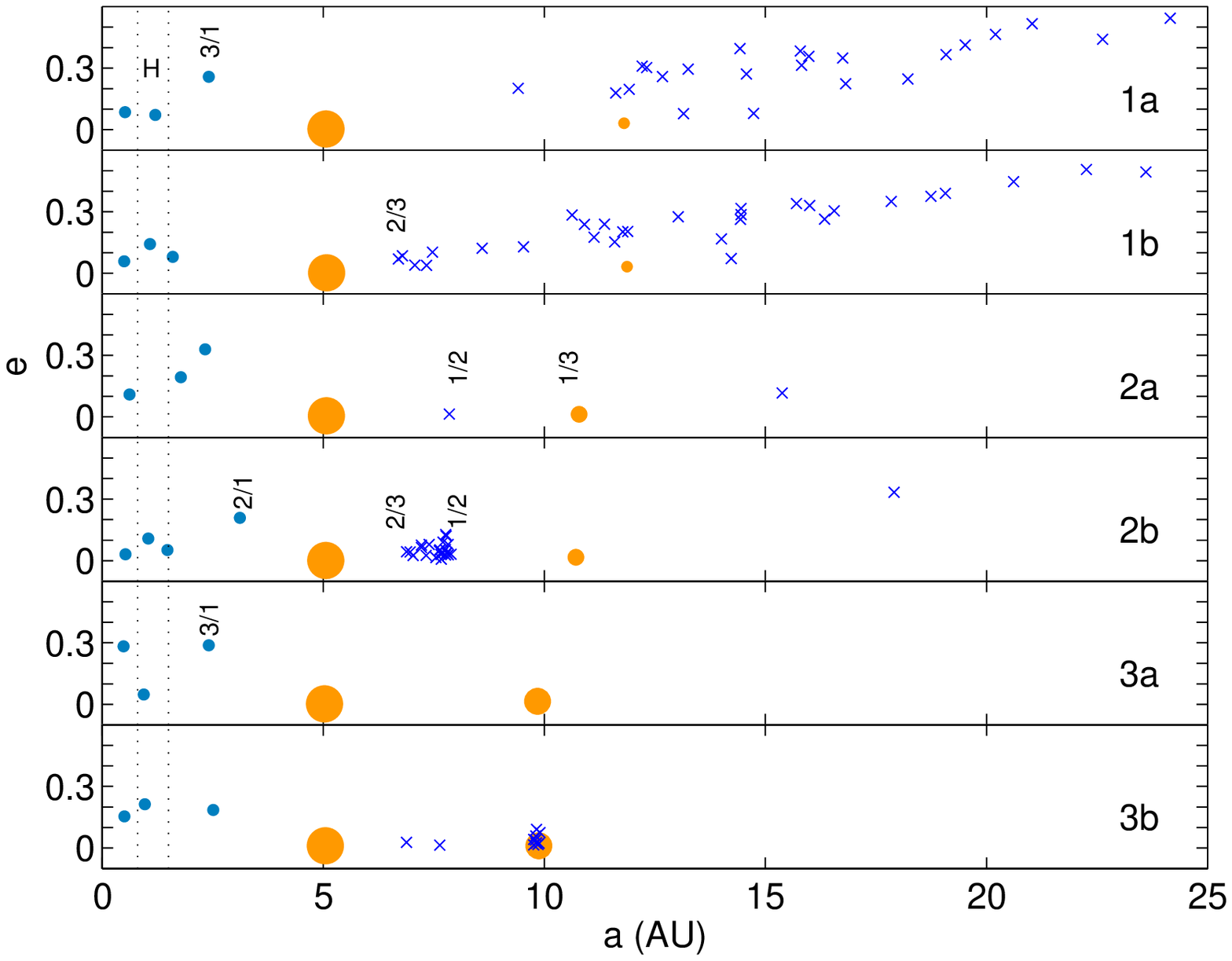} \caption{Results of six simulations.
The $\bullet$ and $\times$ denote formed terrestrial planets and
survival planetesimals. Jupiter and Saturn locate at about 5 AU and
10 AU,  respectively. $H$ zone marked with dotted lines is so called
Habitable Zone in 0.8 - 1.5 AU. Some mean motion resonance locations
with Jupiter are also labeled in the figure. \label{fig5}}
\end{figure}

\begin{figure}
\figurenum{6}
\plotone{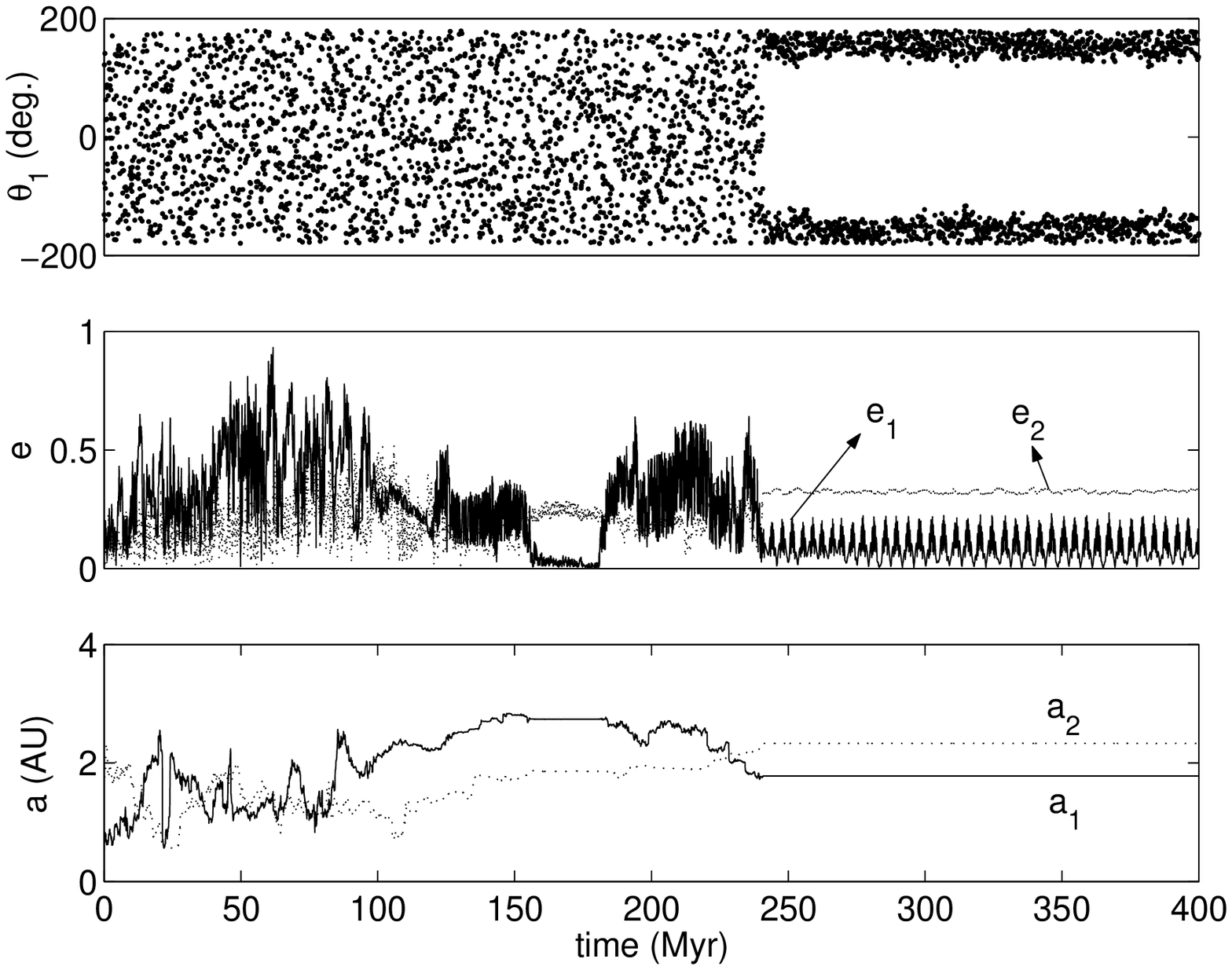}
\caption{Resonant variable
$\theta_1$ of the 3:2 mean motion resonance of two terrestrial
planets and the curve of eccentricities and semi-major axes in
simulation 2a. ($\theta_1=2\lambda_1-3\lambda_2+\varpi_2$, where
$\lambda _{1,2} $ are the mean longitudes and the longitudes of
periapse, the footnotes 1, 2 mean the inner and outer planets
respectively.)\label{fig6}}
\end{figure}

\end{document}